\documentclass[journal]{IEEEtran}
\usepackage{graphics,graphicx,epsfig,amssymb,amstext,amsmath,algorithm,
algorithmic,wrapfig,multirow}
\usepackage{mathtools}
\usepackage{cite}
\usepackage{url}
\usepackage{times}
\usepackage{float}
\usepackage{placeins}
\usepackage{textcomp}
\usepackage[font=small,bf]{caption}
\usepackage{hyperref}
\usepackage{flushend}
\usepackage{xcolor}
\usepackage{tabularx}
\usepackage{bm}
\usepackage{enumerate}
\usepackage{tikz}
\usepackage{enumerate}
\usetikzlibrary{shapes,arrows,patterns}
\usepackage{epstopdf}
\usepackage{amsthm}
\usepackage{subcaption}

\DeclareGraphicsExtensions{.eps}

\def\beq{\begin{equation}}
\def \eeq{\end{equation}}
\def\beqa{\begin{eqnarray} }
\def\eeqa{\end{eqnarray} }
\def\beqan{\begin{eqnarray*}}
\def\eeqan{\end{eqnarray*}}

\def\tm1{t\! - \! 1}
\def\tp1{t\! + \! 1}

\newif\ifconf
\conffalse

\newif\ifonecol
\onecolfalse
\IEEEoverridecommandlockouts

\renewcommand{\footnoterule}{
  \kern -3pt
  \hrule width \columnwidth height 0.5pt
  \kern 3pt
}
\begin{document}
\tikzstyle{sig} = [draw=none, rectangle,pattern=north west lines, pattern color=blue!80, minimum height=4cm, minimum width=0.4cm]
\tikzstyle{sig2} = [draw=none, rectangle,pattern=north west lines, pattern color=red!80, minimum height=4cm, minimum width=0.2cm]

\title{Achieving Ultra-Low Latency in \\ 5G Millimeter Wave Cellular Networks}

\author{
    Russell Ford,~\IEEEmembership{Student Member,~IEEE},
    Menglei Zhang,~\IEEEmembership{Student Member,~IEEE},
    Marco Mezzavilla,~\IEEEmembership{Member,~IEEE},
    Sourjya Dutta,~\IEEEmembership{Student Member,~IEEE},
    Sundeep Rangan,~\IEEEmembership{Fellow,~IEEE},
    Michele Zorzi,~\IEEEmembership{Fellow,~IEEE}
    \thanks{This material is based upon work supported by the National Science
    Foundation under Grants No. 1116589 and 1237821 as well as generous support
    from NYU WIRELESS affiliate memberships.}
    \thanks{
        The authors are with NYU WIRELESS, New York University Polytechnic
        School of Engineering, Brooklyn, NY 11201 USA
        (e-mail: {xxx}@nyu.edu.  Michele Zorzi is with the University
        of Padova, Italy (email: zorzi@dei.unipd.it)}
}

% make the title area
\maketitle

\begin{abstract}
The IMT 2020 requirements of 20 Gbps peak data rate and 1 millisecond latency present significant engineering challenges for the design of 5G cellular systems. Use of the millimeter wave (mmWave) bands above 10 GHz ---
where vast quantities of spectrum are available ---
is a promising 5G candidate that may be able to rise to the occasion.

However, while the mmWave bands can support massive peak data rates,
delivering these data rates on end-to-end service while maintaining
reliability and ultra-low latency performance will require
rethinking all layers of the protocol stack. This papers surveys
some of the challenges and possible solutions for delivering
end-to-end, reliable, ultra-low latency services in mmWave cellular
systems in terms of the
Medium Access Control (MAC) layer,
congestion control and core network architecture.

%The millimeter wave bands cellular systems and have been shown to achieve multi-Gigabit throughput by exploiting the large quantities of spectrum available at higher frequencies along with high-dimensional antenna arrays.
%a novel radio frame and control plane design is needed. Additionally, an efficient Medium Access Control scheme is required to manage spectral resources and meet the QoS constraints of each user.

%In this paper, we propose a frame structure and control plane design for a millimeter wave cellular system that enables ultra-low latency communication and incorporates Hybrid ARQ with Incremental Redundancy (HARQ-IR) capability to improve reliability. We then propose a throughput-optimal Dynamic Time-Domain Duplexing MAC scheduling algorithm that can achieve 1-ms latencies for a feasible set of real-time traffic flows. Finally we demonstrate the latency and throughput performance of the system under realistic traffic scenarios through detailed, full-stack discrete-event simulation in NS-3.
\end{abstract}

\section{Introduction}

Millimeter wave (mmWave) communication is widely considered to be a promising candidate technology for fifth generation (5G)
cellular and next-generation wireless Local Area Networks (LANs).
Already, the wireless industry is investing heavily in developing systems that operate in the millimeter wave bands, which are attractive
because of the vast quantities of virgin spectrum and the spatial degrees of
freedom afforded by very
high-dimensional antenna arrays (thanks to the smaller size of antenna elements at higher frequencies). Regulatory agencies are also beginning to consider defining new licensed and unlicensed bands for commercial use.
Although mmWave radio links are already used in a variety of commercial applications, until recently they were considered impractical for mobile access networks
due to extreme vulnerability to shadowing and poor isotropic propagation loss.
Results from recent measurement campaigns have demonstrated that the limitations of the mmWave channel can indeed be overcome by high-gain smart antennas, meaning that these large swaths of spectrum can now, for the first time, be exploited to provide an order of magnitude or more increase in throughput for mobile devices \cite{RanRapE:14,AkdenizCapacity:14}.

The news of these measurement results and the promise of massive bandwidth could not have come sooner for operators confronting the surging demand for mobile data. Ultra-wideband millimeter wave is also eagerly welcomed by engineers as one means of achieving the 100 Mbps cell edge and 20 Gbps peak rate specified by such bodies as the ITU and FP7 METIS 2020 Project. Although millimeter wave prototypes have already been demonstrated that can transmit at multi-Gigabit rates \cite{gozalvez20155g}, the oft-cited requirements of 1 millisecond \textit{over-the-air} latency and a ``near instantaneous'' user experience are perhaps even more daunting than the need for extremely high throughput. Emerging use cases like mission-critical Machine-Type Communication (MTC) along with the anticipated ``killer apps'' of the \textit{Tactile Internet}, like immersive virtual reality, augmented reality and telesurgery, to name a few, present a need for \textit{ultra-low latency} mobile networks \cite{fettweis2014_tactile}.

However, while the mmWave bands potentially enable ultra-low latency and massive
bandwidths at the \emph{physical} layer, realizing this very high
level of performance
for \emph{end-to-end} services presents enormous challenges.
End users and applications experience latency at all layers of the protocol stack. Hence, many aspects of the way that cellular systems are designed
will need to be reconsidered if the full potential of the mmWave bands
are to be fully realized.
The paper surveys some of the challenges and possible solutions
for delivering high rate, ultra-low latency end-to-end services in
5G cellular systems. The survey will focus on three critical higher-layer
design areas:  (i) low-latency core network architecture;
(ii) flexible MAC layer; and (iii) congestion control.

%
%
%Faced with these challenging requirements, is clear that achieving the performance standards of 5G will necessitate overhauling the cellular Radio Access Network (RAN) and core network. Despite being a highly-optimized system, 4G LTE-Advanced and the Evolved Packet Core (EPC) are simply not up to the task in their current form.
%
%In Section \ref{sec:5g_arch}, we discuss some limitations of the 4G core network and review the state-of-the-art research on new Software Defined Network (SDN) and Mobile Edge Cloud (MEC) architectures, which aim to reduce end-to-end latency and provide the Quality of Experience (QoE) that users will expect from future 5G networks. In Section \ref{sec:mac_aspects}, we consider the design of a Medium Access Control (MAC) protocol and propose a radio frame structure and control plane that can offer sub-millisecond radio link latency. We then continue the discussion of end-to-end latency and address the performance of TCP over the mmWave channel in Section \ref{sec:tcp}. We finish with our overall conclusions and key take-aways in Section \ref{sec:conclusions}.

\section {Core Network Architecture Challenges}
\label{sec:5g_arch}
\begin{figure*}[ht]
\centering
\includegraphics[width=1\textwidth]{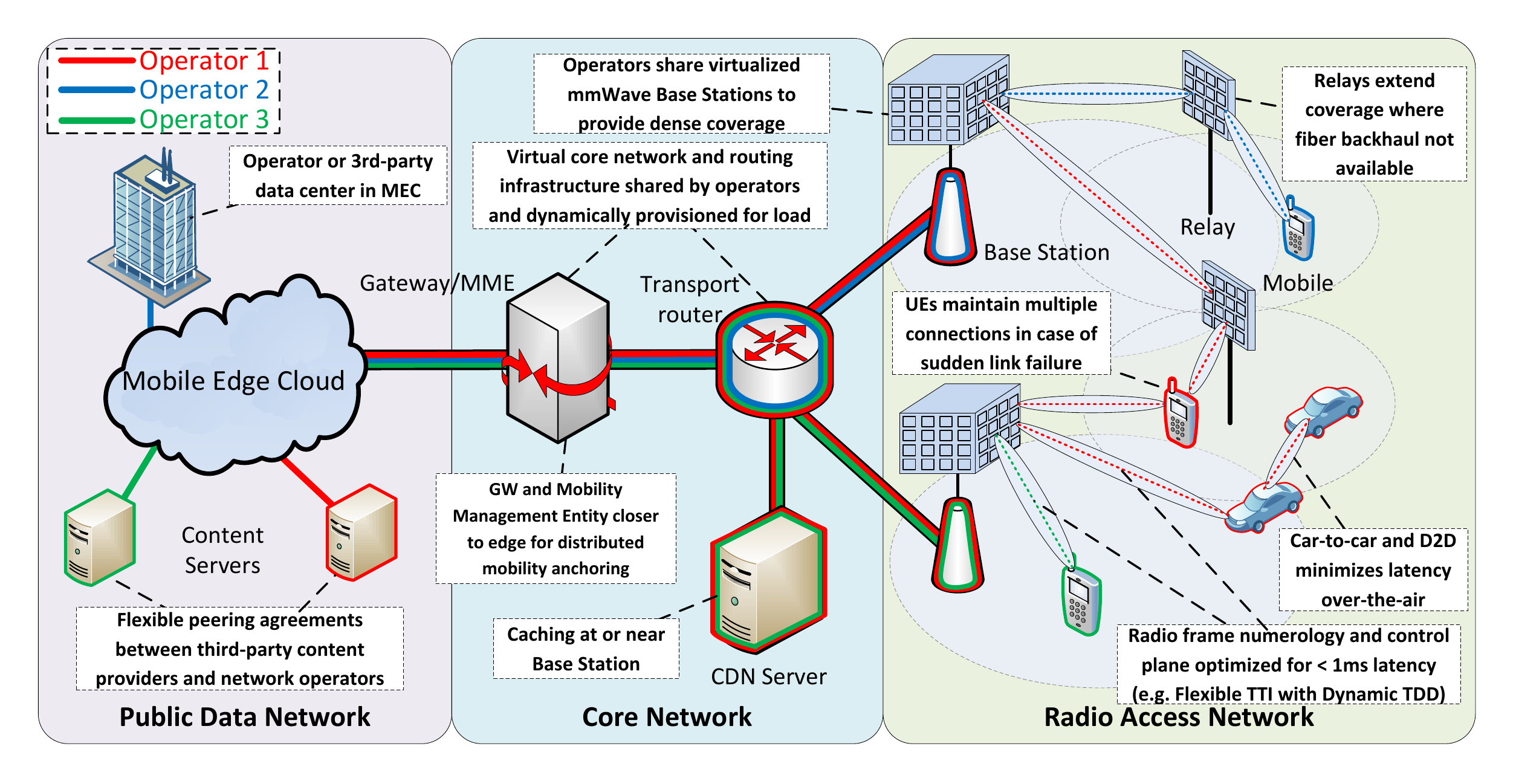}
\caption{Realizing ultra-low latency from an end-to-end perspective will require innovations throughout the network.}
\label{fig:low_latency_net_arch}
\end{figure*}

%The proposed Key Performance Indicators (KPIs) and emerging use cases for 5G call for a comprehensive reworking of mobile network architecture. The mounting requirements for 4G and, soon, 5G services put pressure on both wireless and wireline operators to densify and upgrade their networks, which comes at a considerable cost. The CapEx of real-estate, fiber backhaul and the base station and core network technology itself, along with the OpEx of maintaining these complex networks to deliver optimum performance to subscribers, has resulted in the sustainability of the current \textit{walled-garden} model of mobile operators providing end-to-end network solutions being called into question. Consequently, the burden of building ultra-dense networks and maintaining multiple generations of cellular technology may, perhaps, lead to an era of more pervasive sharing of infrastructure and peering between operators and service providers. Also, with innovations in the 60 GHz space and the possibility of other large bands in the mmWave range being allocated for unlicensed access, the barrier to entry may be lowered allowing new providers to enter the market and offer a more diverse range of services.

To understand the challenges in delivering very low-latency services,
it is useful to begin by considering the typical cellular network architecture
over which data is delivered.
Figure ~\ref{fig:low_latency_net_arch} depicts a simplified Evolved Packet Core
(EPC) network, including the gateway (packet and serving gateway, P-GW/S-GW, in EPC terminology),
which is the IP anchor point for users connecting to the Internet. P-GW/S-GW nodes maintain all of the bearers which tunnel traffic between Internet servers and the User Equipment (UE).

According to the METIS 2020 requirements \cite{popovsk2013eu}, target E2E latency can be below 10 ms for some use cases, which cannot currently be realized in either the 4G core or, as we discuss in the next section, the Radio Access Network. One fundamental bottleneck in the current
Evolved Packet Core (EPC) is the physical distance over which data must be forwarded between the Internet and the user device.
Each gateway node often covers a large geographic region and can be responsible for forwarding and filtering traffic for hundreds of thousands of connections, a task which requires large, high-reliability, costly-to-maintain data centers. Inherently, the geographic distance and transport network hops between the gateway and the end-users incur considerable delay, even before packets reach the Internet.

To deliver very low-latency services, gateway nodes and network points
of attachment will have to be moved closer to the network edge.
Hence, there is  not only a need for the RAN to become more dense to improve coverage but also for core network entities to become more distributed and located closer to the end-user.

Several network technologies are being investigated to create this
more flexible and open network architecture.
In 5G, Software-Defined Networking (SDN) and Network Function Virtualization (NFV) are two recent trends that lend to more distributed topologies, presenting opportunities for lower latency  \cite{Design_5G_CommMag:14}. Additionally, they enable sharing and dynamic provisioning of network resources and functionalities to reduce CapEx and OpEx. SDN enables operators to configure flexible virtual network \textit{slices}, which are overlay networks sharing a common infrastructure. SDN is often combined with NFV, which realizes network functions as separate Virtual Machines (VMs). By decoupling these VMs from the underlying infrastructure, it is possible to dynamically deploy core network entities such as packet gateways and soft switches close to the edge network to adapt to varying load. NFV offers similar benefits from a virtualized RAN, with virtual base stations that can be brought up to increase capacity and torn down when lightly loaded. SDN slices also enable multi-tenant networks where multiple operators coexist and readily share and internetwork base stations, routing and core network nodes toward providing seamless, low-latency handover. The highly unreliable nature of mmWave links may require that UEs maintain connections to multiple base stations, possibly operated by different carriers, making the tighter internetworking and seamless roaming afforded by SDN-based mobility attractive.

Distributed, virtual mobile networks can be combined with content distribution and other services hosted in the edge network to realize what is called the Mobile Edge Cloud. Services, hosted in VMs, can be co-located and dynamically provisioned in edge data centers or in the RAN itself to minimize latency. For instance, a content caching VM may be instantiated at a virtual Base Station to provide real-time navigation data to self-driving cars.

%To compliment such innovations in the core network, the drive for sub-millisecond over-the-air latency also presents the need for new radio stack, frame structure and numerology that is compatible with new physical-layer access technology, such as millimeter wave. We continue in the next section with discussion of a MAC-layer design and radio frame that is suitable for meeting the stringent requirements of the 5G RAN.

\section {MAC Layer Design Issues}
\label{sec:mac_aspects}

\subsection{Challenges}

Current 802.11 wireless LAN systems are easily able to achieve
sub-millisecond airlink latencies.  However, delivering very low latencies
in cellular systems is significantly more challenging.
Cellular systems by nature
must accommodate large numbers of users per cell
and incur significant delay for scheduling and coordinating
transmissions and adjusting to variable channel conditions to maximally
utilize the airlink resources.
Indeed, the current minimum data plane latency in 4G LTE is in the order
of 20~ms, and can be even higher with multiple retransmissions.
Thus, 5G mmWave medium access control (MAC) will need to be redesigned
to reduce latency by at least an order of magnitude.

A key challenge for mmWave systems is that transmissions must be
highly directional to overcome the high isotropic path loss.
Most transceivers, at least in the near future, are likely to use
phased arrays to direct beams.  These arrays can achieve very high directional
gains but are limited to transmitting one user at a time, i.e., via
Time-Division Multiple Access (TDMA) scheduling.
Unfortunately, TDMA can lead to a potentially very poor resource utilization
since the entire bandwidth must be allocated to a single user.
This allocation can be very inefficient for short MAC-layer and higher-layer
control messages.  Moreover, to achieve very low latency and react to
very fast varying channels, control channels
such as scheduling requests and channel quality indicator (CQI) report
will need to have frequent opportunities for transmissions.  These short
control channels will thus incur significant overhead if they cannot be transmitted
efficiently.

%Adaptive beamforming (BF) with multi-element antennas is a key enabler for mmWave communication and, while this technology offers significant gains in terms of power and range, it also introduces new challenges and limitations for designing practical systems. As discussed in \cite{Dutta:15}, the type of beamforming capabilities available to the transmitter and receiver has implications for the MAC Layer frame structure and control signaling. The first generation of mmWave systems, such as the prototypes presented in \cite{PiSysDes:11} and \cite{AGhosh:14}, will likely rely on analog BF where the transmitter focuses its power in a single direction and therefore toward a single receiver (as opposed to omni-directionally). Since communication can only effectively take place between a pair of devices at a time, Time Division Multiple Access (TDMA) is the natural scheme for multi-user mmWave downlink when only analog BF is available \cite{Dutta:15}. However, TDMA with fixed slot lengths or Transmission Time Intervals (TTI) can result in poor resource utilization and latency in traffic scenarios where many small packets must be transmitted to or received from many User Equipment (UE) devices. Such a traffic pattern is typical of the MTC use case, for which we may see hundreds of these low-rate devices. %A fixed-TTI scheme can be especially wasteful for ultra-wideband mmWave systems for which megabytes of data can be transfered within a fraction of the standard 1 millisecond TTI for LTE.

\subsection{Potential Solutions}
\label{sec:frame_structure}
%We first define some basic parameters applicable to the frame structure within our system model, which extends our previous work in \cite{Dutta:15}. We generally adopt the notation used in LTE. Also, in the vein of other proposed mmWave designs (see \cite{PiSysDes:11}), we assume an OFDM waveform in this work. However, we note that the parameters defined in this section can apply to other waveforms like SC-FDMA as well (by simply substituting OFDM ``symbols'' for Single Carrier ``blocks,'' which are defined in \cite{AGhosh:14}).

%Let time be divided into frames, indexed by $i_{f}$, of period $T_f$. Frames are divided into subframes $i_{sf}$ of period $T_{sf}$, which are composed of $N_{sym}$ OFDM symbols of duration $T_{sym}$. Subframes are, in turn, subdivided into variable-length slots $i_{slot}$ of duration $T_{slot}(i_{slot})$ and are made up of $\tau_{slot}(i_{slot})$ OFDM symbols, where $\sum_{i_{slot}} \tau_{slot}(i_{slot}) \le N_{sym}$. Each slot can be assigned to a different user in either Downlink or Uplink mode, with the assignments signaled as part of the control information as described in the next section. Subframes also have special reserved symbols for the Downlink Control (DL-CTRL) and Uplink Control (UL-CTRL) periods, with the remaining symbols available for either DL or UL data.

%We now describe the basic structure of the TDMA frame format.

To deliver very low-latencies at the MAC layer, there are at least three key
modifications one could consider with respect to current 4G LTE OFDM systems:
\begin{itemize}
\item \emph{Short symbol periods:}  Efficient TDMA
transmission of short control messages
requires that one can allocate control transmissions in very short time intervals.
LTE uses OFDM, which enables very simple equalization.
In OFDM, the minimum allocation is one symbol period, which in the current LTE system
is 71.4 $\mu$s (for normal CP). To improve the utilization, several designs have proposed
using much shorter symbol periods, in the order of 4~$\mu$s.  The short OFDM symbol
period can be used for mmWave systems since these systems
are targeting very small cells with low
delay spreads (typically under a few hundreds of nanoseconds).

\item \emph{Flexible TTI:} In current LTE systems, transmissions are sent on a fixed transmission
time interval (TTI) of 1~ms.  With TDMA scheduling, allocating data to any reasonable-sized
fixed TTI would be very inefficient for small packets that would not be able to fully utilize the TTI.
Thus, variable TTI-based TDMA frame structures have been proposed in \cite{Dutta:15} and \cite{kela2015novel}. Also known as Flexible TTI, these schemes allow for slot sizes that can vary according to the length of the packet or Transport Block (TB) to be transmitted and are well-suited for diverse traffic. The flexibility in resource scheduling afforded by a variable TTI system allows both intermittent and bursty traffic with small packets (characteristic of MTC) as well as high throughput flows like streaming and file transfers to be handled efficiently. The concept of flexible TDMA therefore harmonizes with the 5G vision of a unified cellular solution for various applications, devices and use cases. 

%Below we present a variable TTI scheme that is amenable for providing sub-1 ms radio link latency. Here we define radio link latency as the delay between buffering a packet at the transmitter and the reception of the corresponding ACK or Negative ACK (NACK) back at the transmitter, after the transmission has been decoded (successfully or not) by the receiver.

\item \emph{Low-power digital beamforming for control:}
As discussed in \cite{Dutta:15}, the utilization of the control channels can be further
improved by digital beamforming.  Analog beamforming with phased shifters forces
the base station to transmit or receive directionally to only one user at a time
and is a major cause of the poor control channel utilization with wide bandwidths.
In general, digital beamforming, which would allow the base station to
transmit and receive in multiple directions, may not be practical due to high power consumption
of requiring A/D converters on each antenna element. However, control channels generally
require very low SNRs, and thus can be transmitted and received with low resolution
A/D without much loss (since the quantization noise will not be significant).
With very low resolution A/Ds (e.g.\ 2-3 bits per element), it is possible that
one could consider a fully digital transceiver, at least for the control channels,
and this could enable frequency division multiplexing within each OFDM symbol
to further improve the control channel utilization.

\end{itemize}

\begin{figure}
\centering
\begin{subfigure}[b]{0.5\textwidth}
\centering
\includegraphics[trim={0 1cm 0 0}, width=1.0\textwidth]{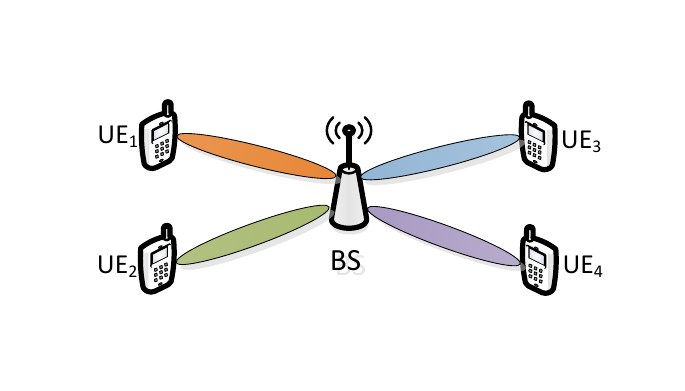}
\caption*{}
\label{fig:mmw_network}
\end{subfigure}
\\
\begin{subfigure}[b]{0.48\textwidth}
\centering
\includegraphics[height=3.2cm,width=1\textwidth]{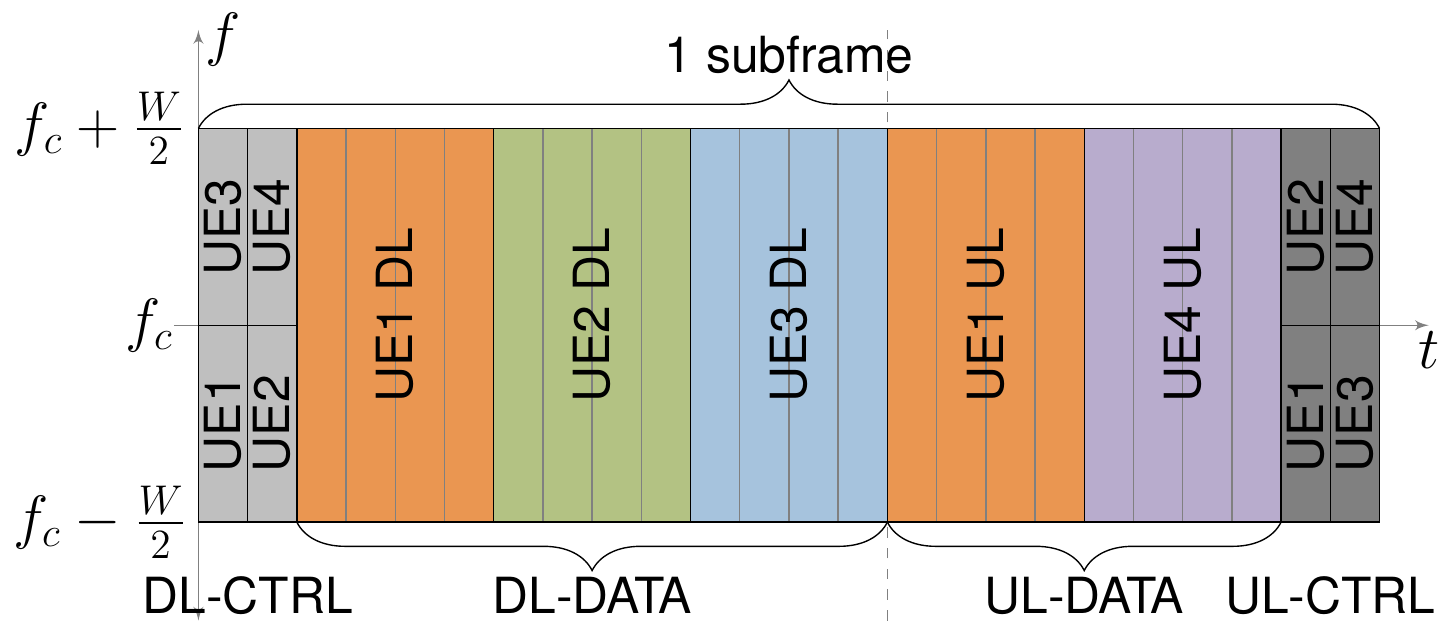}
\caption{Fixed TTI subframe}
\label{fig:fixed_tti_subframe}
\end{subfigure}
\\
\begin{subfigure}[b]{0.48\textwidth}
\centering
\includegraphics[height=3.2cm,width=1\textwidth]{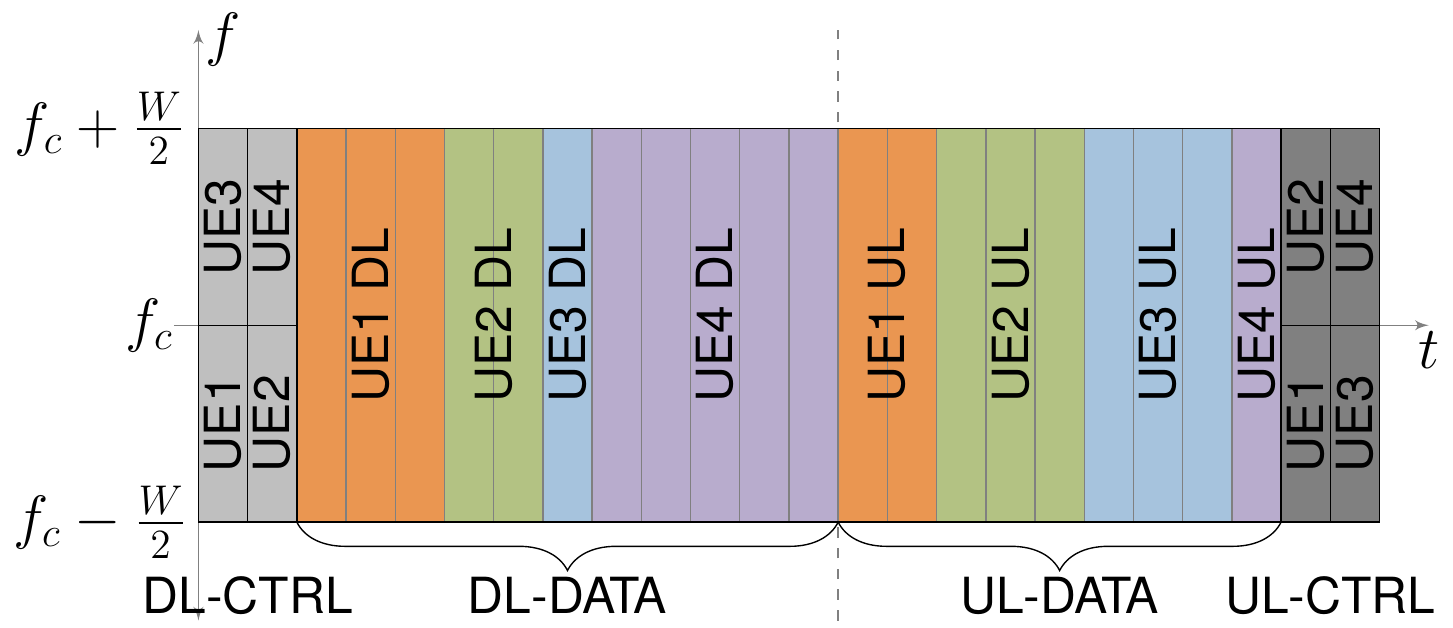}
\caption{Variable/Flexible TTI subframe}
\label{fig:var_tti_subframe}
\end{subfigure}
\caption{Variable and Fixed TTI subframe formats for Dynamic TDD}
\label{fig:frame_structure}
\end{figure}

\subsection{Low-Latency mmWave MAC}

To evaluate the possible achievable latency with a flexible TTI frame
structure, we consider the frame structure in Figure \ref{fig:frame_structure},
similar to the design recently proposed in \cite{Dutta:15} and \cite{kela2015novel}. The key components
are as follows:

\paragraph{Data channel} As previously noted, we consider a system where data transmission relies on analog BF, with the optimal Angle of Departure (AoD) and Angle of Arrival (AoA) selected to maximize SNR between a specific transmitter (TX) - receiver (RX) pair (i.e., one UE and its serving Base Station). Therefore, transmission of data slots is strictly TDMA-based and, as a consequence, the minimum time-domain granularity of resource allocation that can be assigned to a single user in the data period (i.e., the minimum slot length) is 1 OFDM symbol. Although mmWave FDMA and SDMA-based systems are certainly possible and can potentially improve spectral efficiency and system utilization, they require more complex digital or hybrid beamforming transceivers along with additional signal processing, control signaling and MAC complexity.

There must be a small guard time between UL and DL transmissions as well as a transition time during which the beamforming vectors are updated at the transmitter and receiver. To reduce the number of these transitions, which are effectively wasted resources, symbols assigned for a particular user are grouped together so that all DL and UL symbol/slot are contiguously mapped to the DL-DATA and UL-DATA regions, respectively. In other words, we allow one slot (of between 1 and $N_{DATA}$ symbols) per UE per subframe and require each UE's symbols to be contiguous, with all DL symbols preceding the first UL symbol in the subframe. In this way, at most one UL-to-DL guard period is needed per subframe and the number of times the TX and RX must synchronize and align their BF vectors is minimized.

\paragraph{Downlink control channel} The DL-CTRL period occupies the first $N_{DL-CTRL}$ OFDM symbols of each subframe. We require that the location and duration of this region be fixed because the control messages it contains are periodic and must be decoded by all or a subset of UEs at the beginning of the subframe. This allows a UE to decode only a small number of symbols to receive any control messages intended for itself, such as the Downlink Control Information (DCI) indicating the DL and UL assignments in the current or future subframe. After reception of the DL-CTRL, the UE can enter sleep mode to conserve power until its appointed slot time (or for the remainder of the subframe if it finds it has no data slots). The ability to rapidly toggle between active and idle mode conserves battery power for mobile devices, a feature which will be particularly vital for power-intensive mmWave systems \cite{lahet2014achieving}. If the control channel for indicating assignments can be transmitted at any time during the subframe, the UE will have to continually decode symbols even when it is not allocated, which would rapidly drain the device's battery.

For Base Stations that do not support digital BF and must transmit the control channel in TDMA mode only, a minimum of one control symbol per allocated user would be needed. Data could, of course, be multiplexed with the control messages for better utilization of the symbol, however, the UE would still have to blindly decode a number of symbols before finding its own DCI. We can imagine other schemes as well where the control information search space is restricted to a subset of symbol indices for a given UE, however such approaches place limitations on mapping of control and data symbols and so are inherently less flexible.

Therefore, there is a strong case for the BS to support digital BF capability in order to multiplex DL control signals to multiple UEs within a single DL-CTRL symbol, either through Frequency-Division Multiple Access (FDMA) or Spatial-Division Multiple Access (SDMA). As a digital BF transmitter requires separate RF chains and antennas for each stream, the number of UEs that can simultaneously receive the DL-CTRL period depends on the antenna and RF front-end architecture.

%If we need to send control messages to a subset $N_{UE}^\prime$ of users in the DL-CTRL period of a given subframe, the number of OFDM symbols needed is $N_{DL-CTRL} \coloneqq \lceil\frac{N_{UE}^\prime}{K_{BF}}\rceil$, where the maximum number of spatial streams as is denoted $K_{BF}$.

%Furthermore, since the transmit power is split between the $K \le K_{BF}$ subbands belonging to each user, the control messages in this period should be encoded as low-rate BPSK or QPSK in order to be received with sufficient effective SNR for decoding.

%Also, we claim that the maximum spatial streams is the limiting factor on the number of Control Channel Elements (CCEs) capable of being transmitted in one control symbol and not the per-CCE bandwidth, which is reasonable because of the wide bandwidth available and the small size of each control message (the details of which are given in the next section).

%In this work, we assume a scheme where OFDMA is employed in conjunction with multi-user BF to ensure orthogonality of each stream, which lends to simplicity of design at the cost of the improved spectral efficiency provided by SDMA (that is, when multiple user's data can be multiplexed in the same time-frequency resources).

\paragraph{Uplink control channel}
The UL-CTRL period is used for the transmission of periodic control messages from the UEs to the BS. In the design presented here, it is located during the last $N_{UL-CTRL}$ OFDM symbols of the subframe so that it is contiguous with the UL data symbols. While its location is fixed within the subframe, like the DL-CTRL, it can occupy a variable number of symbols depending on the number of users and control messages to be transmitted. Here FDMA or SDMA can be employed along with analog BF performed by each UE to directionally transmit control signals over different subbands or spatial streams. If digital BF capability is not available, the BS can set its phased array antenna for omnidirectional reception, although the reduction in RX antenna gain from doing so will require UL control signals to be encoded at a lower rate (thereby occupying more channel degrees of freedom) in order to attain sufficient effective SNR for being decoded reliably. RX digital beamforming thus has the advantage of reducing uplink control overhead when compared to omnidirectional reception (as demonstrated in \cite{Dutta:15}).

\subsection{Simulating Latency for Small Packets}
\begin{figure}[ht]
\centering
\includegraphics[width=0.5\textwidth]{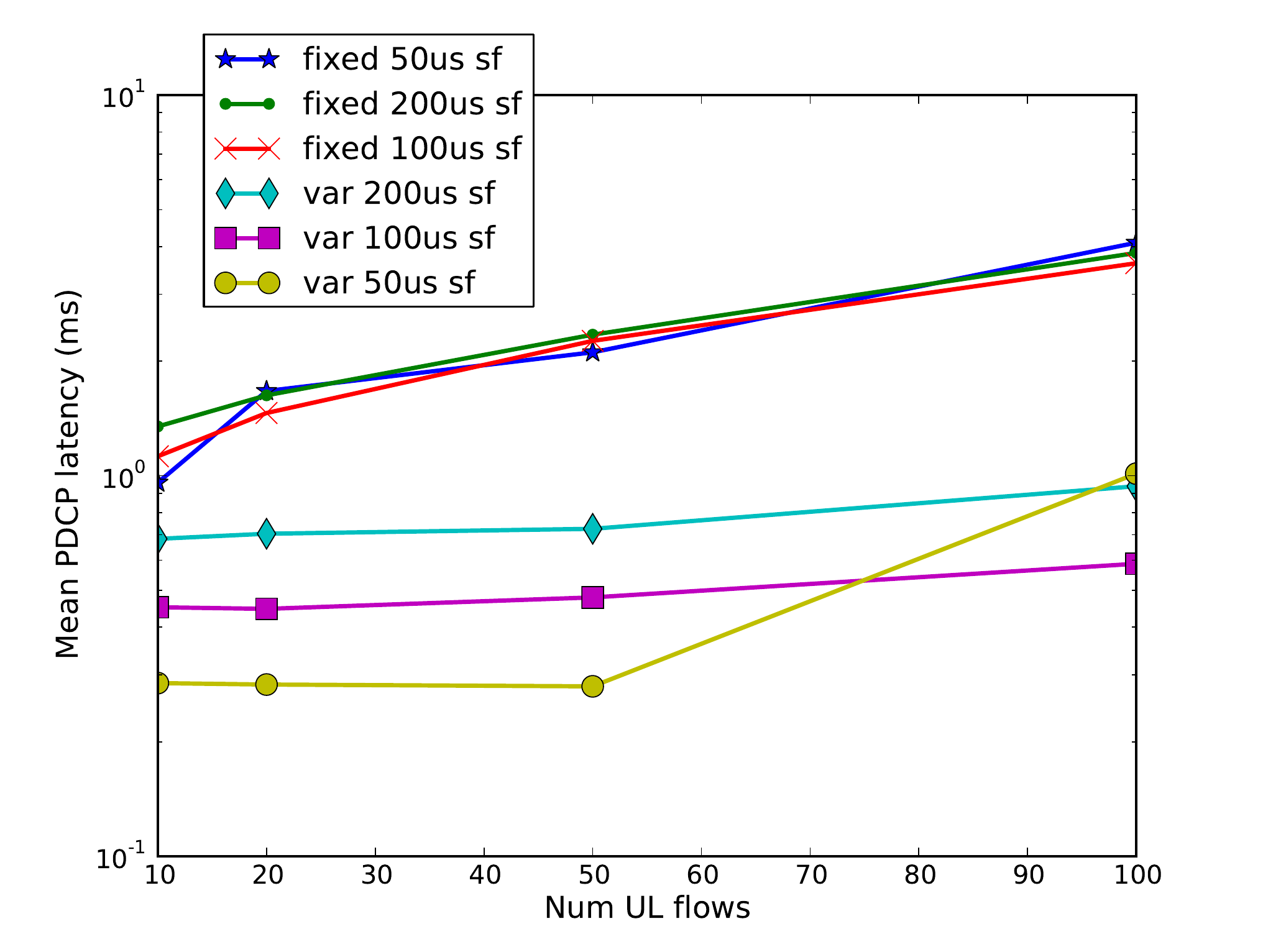}
\caption{Uplink latency for variable vs. fixed TTI subframes vs. number of user flows}
\label{fig:ul_latency}
\end{figure}
To give the reader a better sense of the benefits of variable TTI over fixed TTI subframes, we simulate a 1 GHz TDMA mmWave system, which serves a number of users each with low-rate traffic. Our simulations make use of the ns-3-based full-stack Discrete Event Simulation model for mmWave cellular networks presented in \cite{MezzavillaNs3:15}, with salient parameters provided in Table 1 of \cite{MezzavillaNs3:15}. We model the subframe formats shown in Figure \ref{fig:frame_structure} for subframes periods of 200 \textmu s, 100 \textmu s and 50 \textmu s, consisting of 48, 24 and 12 OFDM symbols, respectively. Each subframe has one fixed DL-CTRL and one UL-CTRL symbol, with the remaining symbols used for DL or UL data slots. For fixed TTI mode, slots must be allocated in multiples of 6 symbols (equivalent to about 25 \textmu s). For variable TTI mode, the scheduler has the freedom to allocate any number of data symbols to each user.

Users are uniformly distributed at distances between 10 and 200 meters from the serving BS and can have either Line-of-Sight (LOS) or Non-Line-of-Sight (NLOS) links, with path loss computed using the model from \cite{AkdenizCapacity:14}. We consider a simple traffic model where each user has both a DL and a UL flow of 100-byte packets being sent at a rate of 100 KB/s. Arriving data packets are scheduled on a first-come, first-serve basis. Furthermore, the slot length required to encode each data PDU is determined by the Modulation and Coding Scheme, which is selected by the scheduler based on the channel state information (as described in \cite{MezzavillaNs3:15}).

Figure \ref{fig:ul_latency} shows the mean uplink radio link latency between the arrival time of packets at the PDCP layer of the UE stack and the time they are delivered to the PDCP layer at the eNB. We see that variable TTI is able to achieve sub-ms latency with hundreds of flows and consistently outperforms fixed TTI, even though the latter uses fairly short 25 \textmu s slots. Even with 10-20 flows there is a distinct improvement, which becomes more dramatic as the network becomes congested with additional users. Also the variable TTI achieves a mean latency under 1 ms even for 100 users, whereas fixed TTI falls below 1 ms only for the 10 user/50 \textmu s subframe case. In the case of 100 \textmu s subframes and 100 flows, we see an improvement of roughly 6x, which follows from the fact that there are up to 6 times the opportunities for users to be scheduled with variable TTI. Also, in the 50-\textmu s subframe case, we observe how, as the number of users increases, the greater ratio of control to data symbols in the 50-\textmu s subframes begins to cause congestion and, in turn, latency.

\section{Congestion Control Considerations}
\label{sec:tcp}

%\subsection{TCP Performance over mmWave Channels}
%\label{sec:tcp}
From an end-to-end point of view,  mmWave communication could create networks
with two features that have thus far never been seen together:
links with massive peak capacity, but capacity that is highly variable.
The massive peak rates arise from the tremendous spectrum available in the mmWave bands 
combined with large numbers of spatial degrees of freedom with high-dimensional antenna arrays.  Indeed, 
recent prototypes have demonstrated multi-Gbps throughput in outdoor environments \cite{gozalvez20155g}.
Simulation and analytic studies \cite{AkdenizCapacity:14,BaiHeath:14} have also predicted capacity gains that are orders of magnitude greater than current cellular systems.
At the same time, the mmWave channel can vary rapidly, making individual links unreliable.
MmWave signals are completely blocked by many common
building materials such as brick and mortar
\cite{KhanPi:11-CommMag,Rappaport:28NYCPenetrationLoss}
and even the human body can cause up to 35~dB of attenuation \cite{LuSCP:12}. As a result, the movement of obstacles and reflectors, or even changes in the orientation of a handset relative to a body or hand, can cause the channel to rapidly appear or disappear.

This combination of features -- massive, but highly variable, bandwidth -- 
presents particular challenges at the \emph{transport layer}, specifically congestion control.
The fundamental role of congestion control is to regulate
the rate at which source packets
are injected into the network to balance two competing
objectives:  (1) to ensure sufficient packets are sent to
utilize the available bandwidth, but (2) to avoid overwhelming the network by sending too many packets, resulting in congestion and affecting other flows in the network.

%To illustrate these challenges, 
%Figure \ref{fig:tcp_perf} shows the performance of a single TCP and UDP flow. 
%\textr{Describe the setting}.  
%The application-layer data rate is fixed at 1 Gbps and a baseline delay from the core and routing network is assumed to be 10 ms. Under good channel conditions, the TCP and UDP flows both send packets at their maximum data rate. However, following a sudden drop in SNR at the 3 second mark, which occurs due to transitioning from a LOS to NLOS path, additional delay is incurred. The sharp loss in capacity results in block errors and, after several HARQ retransmissions of the same process, RLC-AM retransmission takes over. Though this MAC-layer failure incurs only several TTIs of delay initially, the cascading effect for subsequent buffered data results in large spikes in a more than 2x increase in RTT for TCP Reno. Also the rate of TCP drops by over 200 Mbps compared to the UDP flow.

\begin{figure}[ht]
\centering
\includegraphics[width=0.5\textwidth, trim={2cm 6cm 2cm 6cm}]{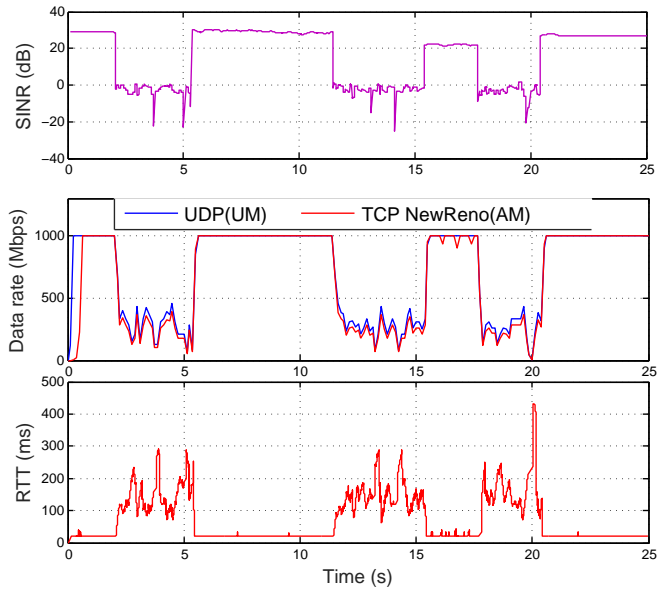}
\caption{Data rate for UDP vs. TCP and TCP RTT for a scenario with rapidly varying SINR}
\label{fig:tcp_perf}
\end{figure}

To illustrate these challenges, Figure \ref{fig:tcp_perf} shows the performance of a single TCP and UDP flow, again simulated using the same model and parameters introduced in \cite{MezzavillaNs3:15}. The application-layer data rate is fixed at 1 Gbps and a baseline delay from the core and routing network is assumed to be 10 ms. Under good channel conditions, the TCP and UDP flows both send packets at their maximum data rate. However, following a sudden drop in SNR at the 3 second mark, which occurs due to transitioning from a LOS to NLOS path, hundreds of milliseconds of additional delay are incurred. This is due to the fact that, when the SINR is high, the TCP client is able to send packets at a high rate and the radio stack is able to service packets at the rate they arrive. However, when the channel capacity is reduced significantly, the buffer becomes severely backlogged as the MAC/PHY-layer can no longer service it at this high rate. Even though the TCP NewReno congestion control algorithm is able to quickly adapt to this sharp loss in capacity, as can be seen in the figure, it is not fast enough to prevent the buffer from becoming backlogged. 

This result raises questions as to the effectiveness of current congestion control and avoidance mechanisms and, perhaps, indicates that some further cross-layer optimization or feedback, facilitated by the network, to the transport or application layer is required to adapt to this high variability.  

\section{Conclusions}
\label{sec:conclusions}

The mmWave bands offer the possibility of a new generation of wide-area
cellular networks with very low latencies and massive bandwidths.  However, 
translating the exciting possibilities of the mmWave spectrum for the physical layer
to end-to-end services will require significant changes
at multiple layers of the protocol stack. This article has identified three 
particular design issues
that need consideration:  (i) changes in the core network to bring data and content
physically closer to the end user; (ii) a flexible MAC layer to enable
low-latency scheduling while still allowing efficient use of the airlink resources;
and (iii) fast adaptive congestion control that handles the rapidly varying nature
of the mmWave channel.  While we have suggested some possible solutions, all these
designs are at a high-level stage and much further will be needed to work out
and evaluate these designs to make these systems a reality. However, if these
technical challenges can be overcome, the potential for next-generation cellular systems
is enormous.

\bibliographystyle{IEEEtran}

\end{document}